# UHV-CVD Growth of High Quality GeSn Using SnCl$_4$: From Growth Optimization to Prototype Devices

P.C. Grant,[1,2,3] W. Dou,[3] B. Alharthi,[3] J.M. Grant,[2,3] H. Tran,[1,3] G. Abernathy,[2,3] A. Mosleh,[4] W. Du,[5] B. Li,[1] M. Mortazavi,[4] H.A. Naseem,[3] and S.-Q. Yu[3,a]

[1]Arktonics LLC., 1339 Pinnacle Dr. Fayetteville, AR, 72701, USA
[2]Microelectronics-Photonics Program, University of Arkansas, Fayetteville, AR, 72701, USA
[3]Department of Electrical Engineering, University of Arkansas, Fayetteville, AR, 72701, USA
[4]Department of Physics and Chemistry, University of Arkansas at Pine Bluff, Pine Bluff, AR, 71601, USA
[5]Department of Electrical Engineering, Wilkes University, Wilkes-Barre, PA, 18766, USA
[a]Electronic mail: syu@uark.edu

**Abstract**: The persistent interest of the epitaxy of group IV alloy GeSn is mainly driven by the demand of efficient light source that could be monolithically integrated on Si for mid-infrared Si photonics. For chemical vapor deposition of GeSn, the exploration of parameter window is difficult from the beginning due to its non-equilibrium growth condition. In this work, we demonstrated the effective pathway to achieve the high quality GeSn with high Sn incorporation. The GeSn films were grown on Ge-buffered Si via ultra-high vacuum chemical vapor deposition using GeH$_4$ and SnCl$_4$ as precursor gasses. The influence of both SnCl$_4$ flow fraction and growth temperature on the Sn incorporation and material quality were investigated. The key to achieve effective Sn incorporation and high material quality is to explore the proper parameter match between SnCl$_4$ supply and growth temperature, which is also called optimized growth regime. The Sn precipitation is significantly suppressed in optimized growth regime, leading to more Sn incorporation into Ge and enhanced material quality. The prototype GeSn photoconductors were fabricated with typical samples, showing the promising devices applications towards mid-infrared optoelectronics.

## I. INTRODUCTION

Progress in group IV photonics over recent years has elevated the GeSn material system to the forefront of photonic integration on Si substrates[1–3]. Demonstrations of true direct bandgap GeSn, LED, and lasers show the growing potential of the GeSn material system[4–9]. Applications based on this technology could flood the photonics market with inexpensive and efficient light emission/detection devices and systems based on these devices.

Growth of GeSn on Ge is difficult due to low solubility (<1%) of Sn in Ge and the instability of α-Sn above 13 °C. In order to grow GeSn material, growth techniques were developed under non-equilibrium growth conditions such as low temperature growth via either MBE [10–16], or CVD [17–24]. The CVD growth of GeSn has been investigated over a decade. Various Sn and Ge precursors in conjunction with carrier gasses were used attempting to achieve high Sn incorporation and high material quality. Early growths were carried out using deuterium-stabilized stannane (SnD$_4$) as the Sn precursor [25], whose high cost and instability drove the motivation to seek other Sn precursors. It has been reported that tin-tetrachloride (SnCl$_4$) is a low cost, stable, and commercially available precursor and the GeSn material growth was initially demonstrated by Vincent at el [26]. On the other hand, various hydride chemistries have also been explored as Ge precursor by Kouvetakis et al [27]. Higher order germanes were commonly used due to their favorable decomposition at low temperatures [28–31]. Our recent progress has allowed for low cost germane (GeH$_4$) to be used for Ge and GeSn growth via UHV-CVD system [19,21,22,32].

Previous reports of GeSn material growth focus on the successful deposition conditions and the final product. While providing useful guidance, they have not provided an effective route to produce high quality GeSn as optimized growth conditions vary with reactor designs and configurations. The growth of GeSn material detailed in the previous growths showed the material properties of the grown films, however no result has yet to demonstrate the entire development pathway to high quality GeSn material. The goal of this work was to detail an optimization methodology for the development of high-quality GeSn material on the Ge buffered Si that is potentially transferable across reactor design and configuration. The results from optimization of GeSn using a home-built UHV-CVD are presented as demonstration of the optimization process. The two dominating factors of growth temperature and $SnCl_4$ flow fraction were extensively explored with wide ranges. Specifically, the GeSn growth starts with the variation of temperature while the $SnCl_4$ flow fraction was kept at the overpressure regime. The cloudy surface of GeSn was observed, which was due to the Sn segregation on the surface. Two distinguished Sn incorporation regions were also determined. Then GeSn growth continues with the decrease of $SnCl_4$ flow fraction while keeping the temperature at the constant, driving the GeSn growth window from $SnCl_4$ overpressure regime to the optimized regime. With the decrease of $SnCl_4$ flow fraction, the mirror-like GeSn surface and effective Sn incorporation were achieved. After that, the growth temperature was further decreased while keeping the $SnCl4$ flow fraction at the optimized region. As a result, the Sn incorporation was further enhanced. Once the high-quality and high-Sn-content GeSn was obtained photoconductor devices were fabricated with infrared imaging also being accomplished.

## II. RESULTS
### A.  Growth Methodology and Sample Description

It has been acknowledged that the combination of Ge hydrides and $SnCl_4$ during the CVD epitaxy results in the severe etching effect, thus limiting the success of GeSn growth. The etching effect originates from the generation of gaseous HCl byproduct during the reaction between Ge hydrides and $SnCl_4$. The overall epitaxy of GeSn is the competition between deposition and etching, both of which heavily depends on the supply of $SnCl_4$ and growth temperature. In order to explore the optimal epitaxy window of GeSn, we designed the experiments by following three logic sequences. First of all, we grew the well-known Low/high temperature two-step Ge films. [33,34] The two-step Ge growth mainly serves two purpose: 1) The low temperature 1st Ge layers helps determining the growth parameters of GeSn as the starting point, including growth pressure, flow rate of precursors, etc; 2) The high quality Ge acts as the buffer layer of GeSn by accommodating the lattice mismatch between GeSn and Si; Second, we demonstrated the success of GeSn growth directly on Si substrate based on the initial exploration of low temperature Ge growth, which has been reported in our precious work [21]. The epitaxy of GeSn film on Si could be easily identified by visual inspection and subsequent material characterization. It also eases the efforts to determine the epitaxy window where the deposition dominates over the etching by visual contrast between GeSn and Si; Third, we grew the high quality GeSn on two-step Ge buffer layer. Based on the knowledge of GeSn growth on Si at the second step, we mainly focus on the optimization of two dominating factors for the effective Sn incorporation: $SnCl_4$ supply and growth temperature.

All material growth work was done using p-type Si (001) 4-inch wafers as substrates. The wafers were processed using standard piranha etch and HF dip prior to growth as described in our previous work [21]. Material growth was carried out in a cold wall UHV-CVD chamber with base pressures below $1\times10^{-9}$ torr. The Growth was accomplished using $GeH_4$, $SnCl_4$ as precursors and Ar as the carrier gas. The Ge buffer layer with approximately 1 µm thickness was grown by low/high temperature two-step growth [33,34]. The 1st 100-nm Ge layer was grown as a seeding layer at 325ºC temperature while the 2nd 900-nm Ge layer was grown at 600ºC. After the growth of Ge buffer, the substrate temperature was cooled down to initiate the GeSn epitaxy. During the GeSn growth the process pressure was fixed at 2 torr. A wide range of growth temperature and $SnCl_4$ flow fraction were explored, as shown in Fig. 1. The flow fractions of $SnCl_4$ was defined as $F_{SnCl_4}/(F_{SnCl_4} + F_{GeH_4} + F_{Ar})$, where the $F_{SnCl_4}$, $F_{GeH_4}$ and $F_{Ar}$ indicate the flow rate of $SnCl_4$, $GeH_4$ and Ar, respectively. Three groups of GeSn was grown on Ge buffer:

**Group 1**: From sample A to F the growth temperature decreases from 325 to 240ºC while the SnCl4 flow fraction was fixed at $2.9\times10^{-3}$. The growth time of samples A, B, C, D, E1, F, G are 30 mins while growth time of sample E2 is 60 mins.

**Group 2**: From sample E2 to I, the $SnCl_4$ flow fraction decreases from $2.9\times10^{-3}$ to $2.3\times10^{-4}$ while the growth temperature was fixed at 270ºC. The corresponding growth time are 60 mins.

**Group 3**: From sample I to K, the growth temperature decreases from 270 to 250ºC while the $SnCl_4$ flow fraction was fixed at $2.3\times10^{-4}$. The corresponding growth time are 60 mins.

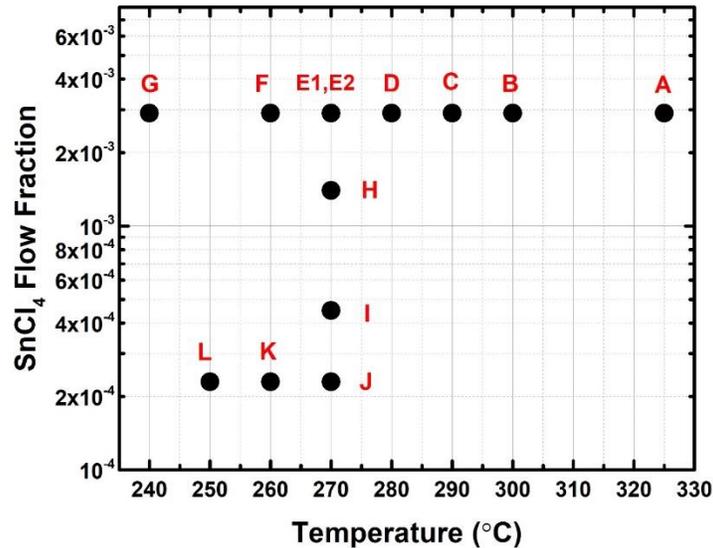

Fig. 1 The growth methodology and sample description of GeSn samples in this paper

The characterization of GeSn films begin as soon as the growth is removed from the chamber by visual inspection. Following the visual inspection, the films are immediately characterized by three techniques: 1) Raman Spectroscopy was used to determine the crystallization of GeSn; 2) spectroscopic Ellipsometry was used for the thickness and absorption coefficient of GeSn; 3) Photoluminescence (PL) was employed for the optical properties of the crystalline GeSn. The PL emission was analyzed by using standard off axis configuration and lock-in techniques with the chopping frequency of 40 Hz. The PL signal was collected using a grating-based spectrometer supplemented with a thermoelectrically cooled PbS detector with cutoff at 3.0 µm. The 1064 nm pulsed laser was firstly used as the pumping source with 45 kHz repetition rate and 6 ns pulsed duration. The laser spot size was measured as 52 µm in diameter. The high carrier injection per pulse ensures the sufficient PL emission from GeSn. The GeSn with strong PL signal was further characterized by using 532 nm continuous wavelength (CW) laser. The film strain, Sn composition and crystallinity was then determined by rocking curves and reciprocal space mapping (RSM) of X-ray diffraction (XRD). The XRD was performed using a Philips X'pert MRD system, which was equipped with a standard four-bounce Ge (220) monochromator and a three bounce (022) channel cut Ge analyzer crystal along with the 1.6 kW Cu Kα1 X-ray tube with vertical line focus. The surface morphology of GeSn was further inspected by scanning electron microscopy (SEM). Film thicknesses and Sn distributions of the typical samples were also cross checked by the cross-sectional transmission electron microscopy (TEM). The TEM images were observed using a Cs corrected Titan 80-300 with a Schottky field emission gun (FEG) operating at 300 kV.

## B. GeSn Growth in the SnCl$_4$ Overpressure Regime

Visual inspection of the wafer after growth can provide useful information about the film surface. Diffusion of light reflecting off the surface is quite noticeable by the human eye and can serve as an indicator of surface roughness. For this work the film surfaces were divided into three categories: i) the sample surface was milky in that visual reflections were not easily visible, ii) the sample surface was hazy such that visual reflections were easily visible but not clear, and iii) the sample surface was clear or mirror-like finish. These categories correspond to the size of Sn droplets seen on the surface with milky have droplets up to 3 μm, hazy films had less than 1 μm droplets, and mirror films had no detectable droplets. The droplet density on the milky and hazy surfaces were <0.5 and >1 droplet per square micrometer, respectively. Visible and SEM images from these classifications are given in section C. Successive growths were accomplished to verify the temperature dependence on Sn incorporation [34]. These temperatures were varied from our starting point determined by the growth of previous materials shown in Reference 22. In this step of the optimization process, all the film surfaces were milky as Sn droplets dominate the surface of the film. Room temperature PL of the first group of films is shown in Fig. 2.

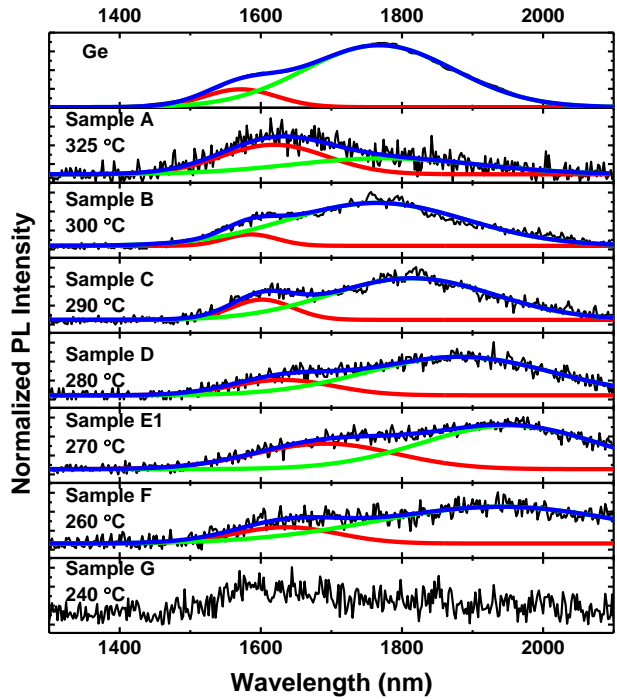

Fig. 2 PL spectra from the growth temperature sweep. The spectra are normalized and stacked to easily show the shifting peak. The dashed lines indicate peak position in the PL spectra.

In Fig. 2, two peaks are evident in the PL spectra below the 325 °C growth temperature. The PL signal is weak and very noisy for all the spectra excluding the Ge bulk reference. The long wavelength peak shifts with reduction in growth temperature, however below 260 °C growth temperature the PL peak disappears. The extension of the wavelength indicates higher Sn incorporation thus confirming the behavior matches what is expected. The short wavelength peak did not shift with decreased growth temperature. This is attributed to the signal from the Ge buffer initially however, the PL wavelength is shifted beyond the buffer emission. This suggests that a separate layer of low composition Sn may be present and is unaffected by growth temperature.

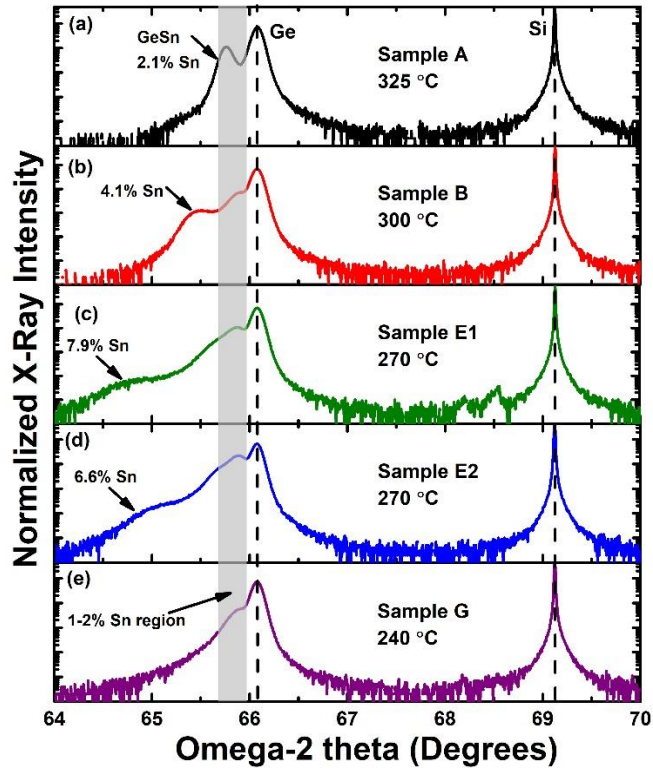

Fig. 3  Selected XRD from the GeSn growth temperature sweep.  Arrows indicate the growth temperature dependent region of GeSn while the shaded area corresponds to a 1-2% Sn layer.

The XRD shown in Fig. 3 shows diffraction peaks for the entire structure. The peak ~ 69° is the Si substrate peak, the peak at 66° corresponds to the Ge buffer peak with GeSn peaks being shifted to angles below 66°. The shifts of GeSn peaks to lower angles with reduction of growth temperature agrees with the extension of PL peak wavelength, indicating increased Sn incorporation. The 325°C growth temperature showed only two layers present in the material however, below 325°C another peak appears and remains throughout the rest of the temperature reduction. This peak does not change with growth temperature reduction. The shift in the low Sn peak, shown in Fig. 3 in the shaded area, from Ge suggests the layer is only 1-2% Sn. At 325 °C the GeSn peak borders along the shaded region and maintains intensity and linewidth like the Ge buffer. Upon reducing the growth temperature below 325 °C, the high Sn peak reduces in intensity and broadens to only a shoulder by 270 °C and is no longer visible at 240 °C. Further investigation of this phenomenon was done to determine the effect of this level of $SnCl_4$ had on the growth of GeSn. Selected samples from the first batch were investigated further to better understand the low

Sn peak seen in XRD. To further investigate this peak XRD-RSM was accomplished as well as TEM. The characterization of a selected sample that is representative of all samples measured is shown in Fig. 4.

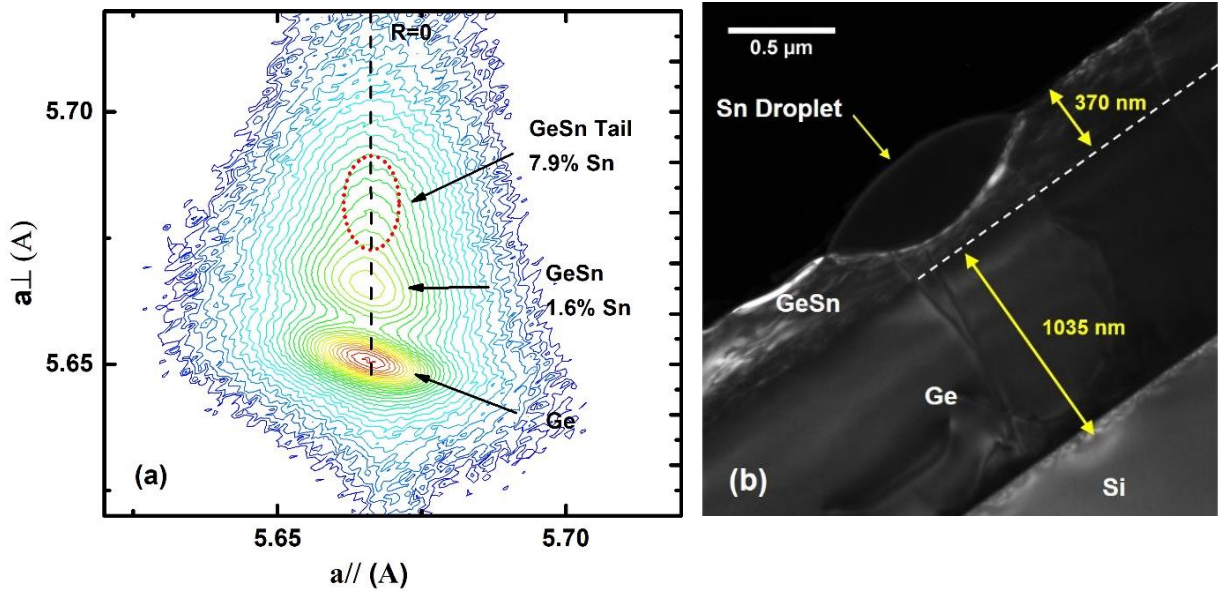

Fig. 4: a) XRD RSM of Sample E1 grown at 270 °C for 30 min. b) Dark field TEM image8 of an area with a Sn droplet.

The XRD RSM of sample E1, in Figure 4a), shows three areas, 1) a Ge buffer peak displaying tensile strain, that was measured at 0.02%, 2) a psuedomorphic layer peak at low composition, (1% Sn), 3) peak from some relaxation in a subsequent layer that corresponds to the growth temperature dependent composition. Below the 300 °C growth temperature, only the low Sn composition layer could be resolved. The samples were then imaged using TEM to see the structure. It can be seen in Fig. 4b, the Ge buffer thickness of 1035 nm is near the 1 µm designed buffer thickness demonstrating good growth control. The TEM image in Fig. 4b also shows a large (>1 µm) Sn droplet on the surface.

PL experiments on these films consisted of two peaks in the spectra with a low Sn peak and a high Sn peak. XRD rocking curve characterization showed that the layer consisted of a low Sn layer with a high Sn tail supporting that two different compositions made up the films. One possible explanation for these results is that the bulk of the film has 1% Sn content, which is covered with Sn droplets and a small amount of Sn has inter-diffused into the layer near the droplet. The RSM clearly shows the 1% Sn layer with some extension that could correspond to the high Sn tail from the rocking curves and the TEM clearly shows the pure Sn droplet on the surface. These pieces of evidence further support the high Sn peak in the PL and rocking curves is localized near the Sn droplet.

## C.  Reduction in Sn incorporation for high quality growth

To fully understand the effect of $SnCl_4$ in the growth of GeSn, the ratio of $SnCl_4$ gas flow ratios was calculated showed that the first batch was grown under a $SnCl_4$ flow fraction of $2.9\times 10^{-3}$. This value is two orders of magnitude above that which has been reported in the growth of high quality GeSn [24]. Therefore, a series of growths were accomplished to reduce the $SnCl_4$ gas supply fraction. These growths were accomplished using the same temperature, pressure, and time (sample H, I, and J). The precursor and carrier gas flow rates were adjusted to reduce the SnCl$_4$ partial pressure. In this process, the SnCl$_4$ supply ratios were calculated for each step and reduced the SnCl$_4$ precursor flow to the optimum level for high quality GeSn material growth. The SnCl$_4$ gas supply ratio was reduced by an order of magnitude from the starting condition to a condition that produced a clear mirror like surface. The initial growth condition showed Sn surface segregation and by SnCl$_4$ reduction resulted in the elimination of surface Sn and clear films. The largest disadvantage of the Sn supply reduction is that the Ge supply was also reduced from 25% to 7% of the gas mixture resulting in thinner films. The same characterization philosophy was applied to the grown film with visual inspection occurring upon removal from the growth chamber. Optical characterizations including PL and XRD were also accomplished. Visual imaging, SEM, imaging, and PL results are contained in Figure 5.

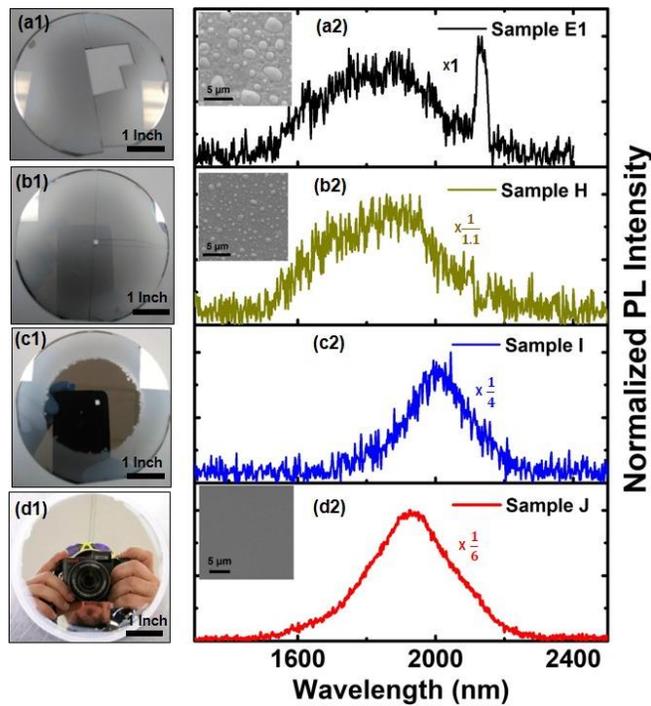

Fig. 5. The visual images of samples E1, H, I and J were shown in (a1), (b1), (c1) and (d1), respectively. From sample E1 to J the flow fraction of SnCl$_4$ decreases from $2.9 \times 10^{-3}$ to $2.3 \times 10^{-4}$ while the temperature was fixed at 270ºC. The room-temperature PLs of samples E1, H, I and J were drawn in (a2), (b2), (c2) and (d2), respectively. The inset: The plan view of SEM images exhibits the cloudy surface (sample E1), hazy surface (sample H) and cloudy Surface (sample J), respectively.

Directly imaging the sample surface shown in Fig. 5 (a1)-(d1) gave evidence of successful Sn reduction. Pieces were taken of Samples E2, H, I, and J, for other measurements before imaging was accomplished. It can be seen in the initial state the sample surface (E2) was milky and sample H wafer surface is still hazy with surface Sn segregation covering the entire wafer surface. However, by sample I the wafer center is clear with only the outer edge being hazy with surface Sn. Sample J shows mirror-like sample surface across the full wafer. This shows that there is a window for the growth with the upper limit on SnCl$_4$ molar flow fraction from sample I is 4.5 x $10^{-4}$. Sample J shows clearer sample surface so the upper limit SnCl$_4$ molar flow fraction is just set to show where the surface Sn is removed, and the wafer surface begins to clear. This does not verify that all the Sn is removed from the surface.

More evidence of the effect of reducing the $SnCl_4$ overpressure can be seen in the surface of the grown samples. It has been shown that $SnCl_4$ overpressure creates droplets on the surface. Visually, as the $SnCl_4$ molar flow fraction reduces, the cloudiness of the sample surface clears up and sample optical quality improves. A more in-depth look at the sample surface was accomplished using a scanning electron microscope (SEM). Figure 5 (a2)-(d2) shows the SEM imaging of the sample surface of selected samples with separate levels of cloudiness.

In Figure 5 (a2), it is shown that the droplet sizes range up to 3 µm in diameter. Another interesting feature of the surface in Figure 5 (a2) is that the surface appears to have pits in the surface where Sn droplets looked to have been located. It is thought that the agglomeration of surface Sn attracts neighboring Sn forming the large clumps on the surface and leaving behind the pits. The hazy surface shown in 5 (c2) shows an accumulation of Sn droplets on the surface much like the surface from the initial state. However, the size of the droplet has decreased in size to be < 1 µm for the largest of droplets. It can also be seen in this figure that the pits left over have also reduced in size, supporting that it is left behind after Sn agglomerates on the sample surface. Figure 5 (d2) shows a clear surface with only a single droplet to be seen in the image. To further gauge improvement of optical quality, PL on the samples was completed and shown in Fig. 5 (a)-(d).

In Fig. 5 (a) and 5 (b), the first two samples (E2 and H) have PL signal have similar intensity and line shape indicating very little change in the quality of the grown films. The extra peak seen in Fig. 5 (a) from sample E2 is due to the second laser harmonic from the 1064 nm pumping laser. Sample I in Fig. 5 (c) showed 4 times increase over sample E2 in PL intensity and shifting to increased wavelengths. This suggests increased incorporation with the Sn surface reduction and higher optical quality of the sample. Sample J in Fig 5 (d) shows a 6 time increase in PL intensity over that of sample E2, indicating higher optical quality than sample I, however the PL peak is shifted back toward shorter wavelengths. The film thickness from sample I is more than that of sample J. The thinner GeSn film should have increased compressive strain which would blue shift the wavelength. Another contribution to the shifting could be minor compositional changes in the growth.

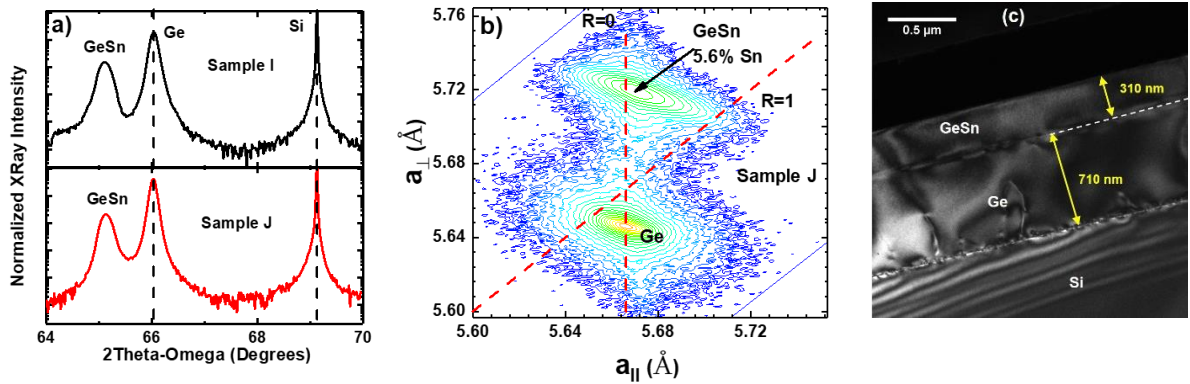

Fig. 6 (a) XRD rocking curves of Sn reduction Tests (samples I and J). The GeSn film thickness of 230 and 160 nm for sample I and J, respectively was derived using spectroscopic ellipsometry. (not shown); (b) XRD-RSM for sample J. Dashed lines show the relaxation when R=0 or pseudomorphic growth, while R=1 is for relaxed growth; (c) Dark field TEM image of sample J. No Sn droplets were observed on the GeSn surface.

The XRD rocking curves contained in Figure 6 (a), show similar peak position for each sample regardless of the flow fraction of SnCl$_4$ or GeH$_4$. Recall from Section B that the shoulder for the sample E2 which is the initial state, had a shoulder in the XRD located at the same angle as samples I and J. No shifting of the XRD with changing of the precursor supply rates shows that temperature is the dominant mechanism that determines Sn incorporation when starting from an initial strain state. The change of XRD peak position is <0.1 degrees suggesting the Sn composition between samples varies less than 1% and is most likely attributed to natural process variation in the growth chamber. More intense peaks from sample I and J and the disappearance of the 1% peak show better material quality and suggests that the Sn has incorporated into the growing film instead of agglomerating on the surface. Composition of the film was determined from XRD RSM shown in Fig. 6 (b), with the GeSn film comprising 5.6% Sn and 11% relaxation. The Ge buffer was found to be under 0.17% tensile strain. The dark field TEM image of sample J was shown in Fig. 6 (c), in which no Sn segregation was observed on the surface. The GeSn film also exhibits no extended threading dislocation penetrating through the film. Improvement in the growth conditions allowing for high-quality growth sets the stage for continued GeSn growth to further increase the Sn incorporation.

## D. More Sn enhancement by lowering temperature

The GeSn films with effective Sn incorporation and mirror-like surface have been achieved by decreasing the Sn dilution ratio down to the gas flow limit of mass flow controller. In order to further increase the Sn incorporation, the growth temperature was designed to decrease while maintaining the Sn dilution ratio at $2.3\times10^{-4}$. By the visual inspection after GeSn epitaxy, the surface morphology of GeSn film changes with the decrease of temperature from the growth temperature from 270 to 250 ºC. As shown in Fig. 7 (b) inset, the majority of GeSn surface is mirror-like without the significant Sn droplet segregating on the surface at the temperature of 270 ºC. When the temperature drops to 260 °C, the hazing area starts growing from the edge to center as the ring on the surface while the GeSn surface at the center is clear as shown in Fig. 7 (c) inset. The hazing area is attributed to the aggregated Sn droplet on the GeSn surface. As the temperature continues decreases to 250 °C, the hazing area continues growing and covers the majority of the surface, leaving the clear surface only sparsely at the center of the surface as shown in Fig. 7 (d) inset.

The room-temperature PL spectra was conducted in order to study the optical properties of GeSn films. As shown in Fig. 7, the PL spectrum of commercialized GeSn sample was drawn as the reference as the comparison with our CVD-grown GeSn samples from J to L. The GeSn reference sample was grown on Ge buffered Si substrate by using the ASM Epsilon® 2000 Plus reduced pressure CVD (RPCVD) system with commercially available precursors of GeH$_4$ and SnCl$_4$. The film thickness of GeSn film is 83 nm. The corresponding Sn composition and compressive strain of commercialized sample was 5% and -0.52%, respectively. On the PL spectra of Fig. 7, the pumping power of 1064 nm laser for PL emission was 300mW and other PL setup conditions between GeSn reference and samples J to K were kept the same. From Fig. 7 (a) to (d) only direct bandgap emission were observed for all the samples while the indirect bandgap emission was not detected. The inefficient indirect bandgap recombination is suppressed by the fast non-radiative recombination induced by defects. From Fig. 7 (a) of GeSn reference, the wavelength of PL peak is 1916 nm. From Fig. 7 (b) to (d) of sample J to L, the PL peak wavelength shifts from 1924 to 2072 nm when the temperature decreases from 270 to 250°C. The pronounced red shift from sample J to K suggested the increased Sn incorporation into Ge matrix when decreasing the temperature. Comparing GeSn reference with our grown GeSn, the wavelength of PL is similar. The FWHMs of reference sample and sample J are 88.5meV (264nm), 96meV (288nm), respectively. The comparison study suggests the promise of our grown GeSn samples in the optical properties close to the reference GeSn samples. The 532 nm continuous-wave (CW) pumping laser was also used for the PL characterization of sample K, as shown in Fig. 7 (c). The pumping power of 532 nm laser is 500 mW and the beam size is 100 μm in diameter. The PL emission of 532 nm laser has the wavelength peak of 2042 nm, which is 118 nm longer than that of 1064 nm pulsed laser. This is due to the pronounced blue wavelength shift of 1064 nm laser pumping introduced by typical band-filling effects as the peak power density per pulse of pumping laser for 1064 nm lasing was calculated as four orders of magnitude higher than that of 532 CW laser. The PL comparison study further confirm the significant improvement of optical performance of our grown GeSn samples.

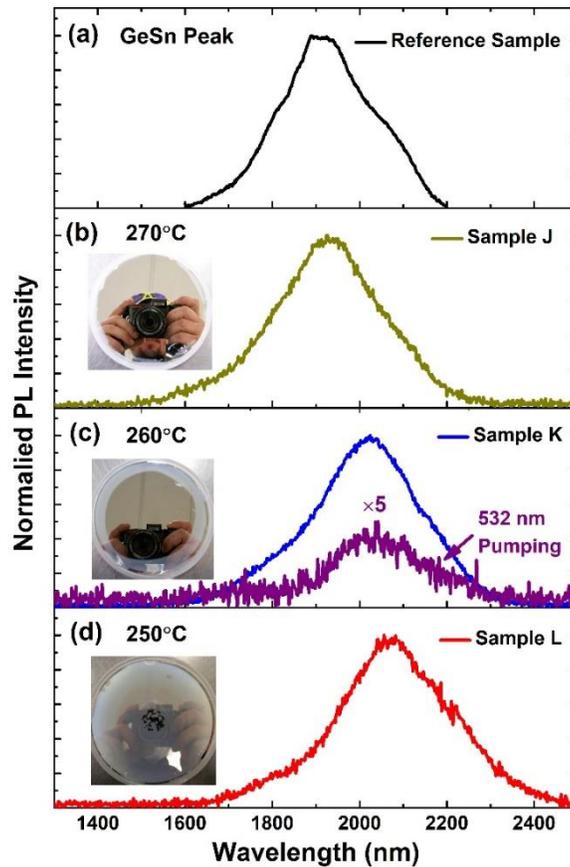

Fig. 7. The room-temperature PL spectra of (a) reference sample, (b) sample J, (c) sample K and (d) sample L, respectively. The 1064 pulsed laser was used as the pumping source for reference sample and samples J, K and L while the 532 nm continuous wave laser was used for sample K as comparison. The samples J, K and L were grown at the growth temperature of 270, 280 and 250 ºC, respectively. The inset: visual images of sample J, K and L.

The XRD technique was conducted in order to investigate the Sn incorporation and compressive strain of GeSn. The rocking curve of commercialized GeSn was drawn in Fig. 8 as the reference. The peaks of Si substrate, Ge buffer and GeSn film were clearly resolved on the rocking curves of GeSn reference and samples J, K and L. From Fig. 8 (a), the GeSn peak of reference is located at 65.08°. From Fig. 8 (b) to (d), the GeSn peaks of sample J, K and L are located at 65.14, 64.82, 64.66°, respectively. The shift towards the lower angles from sample J to L indicates the increase of Sn incorporation when temperature drops from 270 to 250°C, which is consistent with the PL analysis. Comparing the GeSn reference and sample J, the GeSn peaks of reference (65.08°) are close to the GeSn peak of sample J (65.14°). The FWHMs of GeSn reference and sample J and was Gaussian fitted as 0.19° and 0.20°, suggesting that the material quality of our grown GeSn is getting close to that of the GeSn reference. It is worth mentioning that at the growth temperature of 250°C a new peak between the main peaks of GeSn and Ge emerges, which is attributed to the GeSn peak with ~1% Sn composition. As the Sn droplets continue growing over the surface at the growth temperature of 250°C, the GeSn with ~ 1% Sn composition reappears, which is in agreement with the analysis in section B.

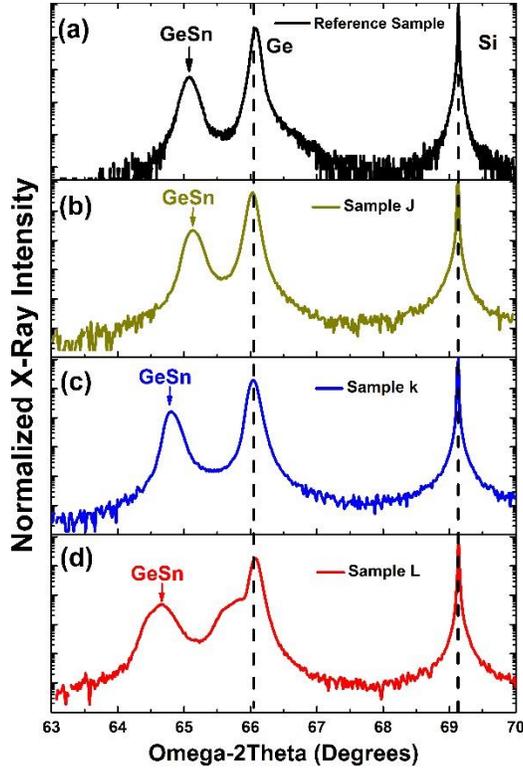

Fig.8. The XRD rocking curves of (a) reference sample, (b) sample K, (c) sample K and (d) sample L, respectively. The Si, Ge and GeSn peaks were marked in XRD rocking curves, respectively.

### E. Prototype GeSn photoconductors

To demonstrate the fully capability of GeSn, photoconductors were fabricated from selected films. Characterization including, I-V, responsivity (not shown), calculated detectivity and infrared imaging were accomplished with a selected result shown in Fig. 9.

The GeSn films were fabricated into photoconductor devices such as the one shown in Fig. 9a. Photoconductors were chosen due simplicity of the device and ease of fabrication. Interdigitated structures were fabricated to allow higher collection efficiency over that of coplanar devices, inset of Fig. 9a. The I-V shown in Fig. 9a consists of both 77 and 300 K measurement. At each temperature the I-V shows linear I-V characteristics suggesting good ohmic contacts. The dark current at 300 K is 40 mA but reduces to below 10 mA at 77 K.

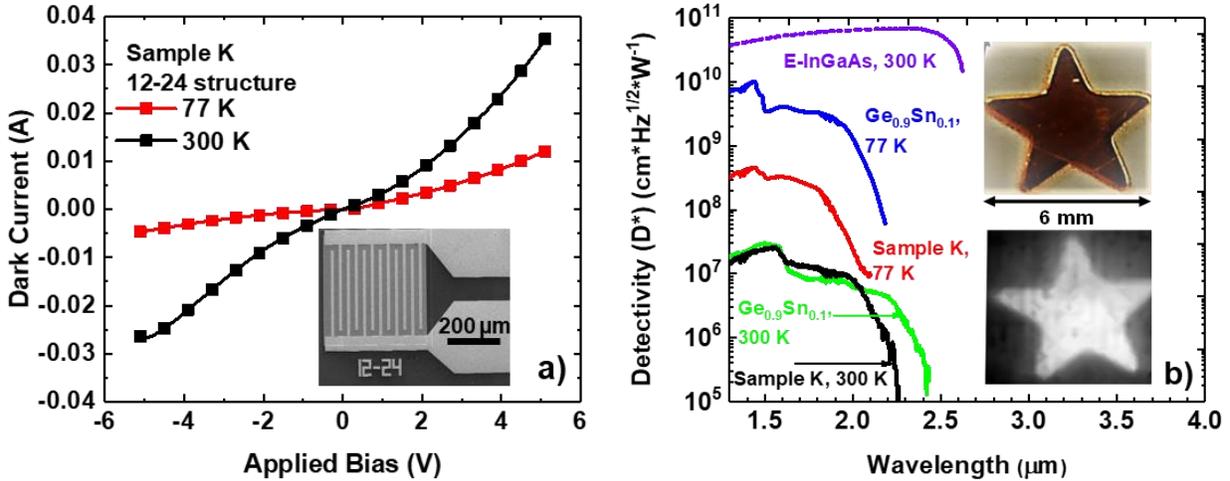

Fig. 9: a) I-V curves for interdigitated GeSn photoconductors. Inset shows measured device; b) Calculated D* for the GeSn photoconductors. Upper inset is a visible image of the object. Lower inset is infrared image generated using the GeSn photoconductor.

The specific detectivity (D*) was calculated for the fabricated GeSn photoconductor and plotted, in Fig. 9b, against commercial extended InGaAs detectors, and a commercially grown 10% GeSn photoconductor that has been previously reported. [35] It can be seen that at 300 K the D* of sample K has similar detectivity to the commercial GeSn with shorter cutoff wavelength, 2.25 and 2.5 μm respectively. However, at 77 the 10% GeSn outperforms the sample K by just over an order of magnitude showing there is still room to improve the growth condition. In Fig. 9b insets, a visible image (upper inset) and an infrared image of an object (star) was demonstrated using the photoconductor fabricated from sample K. Imaging was accomplished a using a single photoconductor moved in an array pattern to generate the image. White light was used to illuminate the 6 mm star and a 1600 nm long pass filter used to filter out non-infrared light from the photodiode. Generating an image from the grown films shows the potential of GeSn.

## III. SUMMARY, DISCUSSION, AND CONCLUSIONS

The experimental results discussed in this work indicates that during GeSn epitaxy the $SnCl_4$ flow fraction has the supply limit for each growth temperature. The high growth temperature corresponds to the high supply limit of $SnCl_4$ and vice versa. Exceeding the $SnCl_4$ supply limit leads to the Sn precipitation on the surface and the deterioration of GeSn material. The key to achieve optimized growth regime in which high quality film and high Sn incorporation is achievd through proper parameter match between Sn flow fraction and growth temperature. In optimized regime, the Sn precipitation on the surface is significantly suppressed and Sn atoms are effectively incorporated into Ge lattice matrix.

For Group 1 samples, the GeSn growth is in the overpressure regime with variable growth temperatures. The excess Sn will segregate on the surface and develop as the droplets due to their low surface free energy. The subsequent Sn atoms supplied by the thermal decomposition of $SnCl_4$ tend to be attracted and absorbed by the high-volume Sn droplets instead of being buried by Ge atoms. The Sn droplets continues growing on the surface via the Ostwald ripening mechanism, [36] thus suppressing the effective Sn incorporation into Ge. In this case, only 1-2% Sn atoms could be incorporated into the Ge lattice matrix, which is close to the Sn solubility (1%) into Ge at the equilibrium condition. High Sn incorporation only occur in the vicinity of Sn-rich GeSn surface with limited material quality.

For group 2 samples, the $SnCl_4$ flow fraction was gradually reduced while keeping the growth temperature at constant. With the decrease of Sn supply, the sizes of Sn droplets on the surface become smaller. Eventually the GeSn growth enters into the optimized growth regime, where the flow fraction of $SnCl_4$ and growth temperature reach a good match. Therefore, no appreciable Sn droplet develop on the surface, leading to more Sn atoms buried into the Ge. Meanwhile the material quality of GeSn was significantly enhanced.

For group 3 samples, the $SnCl_4$ flow fraction was fixed while decreasing the growth temperature. It has been reported that at the optimized growth regime the decrease of growth temperature leads to the increase of Sn incorporation, which was further verified in this work. However, the continuous decrease of temperature break the good match between $SnCl_4$ supply and growth temperature and shifts the growth condition into $SnCl_4$ overpressure regime. As a result, the Sn atoms start segregating on the surface, which again jeopardizes the effective Sn incorporation and material quality.

In conclusion, this paper has discussed the method pursued to grow high quality GeSn on Ge buffered Si. The study started in a growth regime in which $SnCl_4$ overpressure dominated the overall growth of the GeSn material. The detrimental effects of $SnCl_4$ overpressure lead to reduction of the $SnCl_4$ molar flow fraction. This reduction in $SnCl_4$ molar flow fraction provided better growth conditions to allow Sn to incorporate into the growing film and less agglomerating on the sample surface. Optimal growth conditions were achieved for the 270 °C growth temperature, producing high quality GeSn with mirror-like surface. Furthermore, temperature reduction below 270 °C resulted in increasing surface Sn with decreasing $GeH_4$ breakdown. GeSn photoconductors were fabricated out of the first and second batch of GeSn growths.